# Technology and Regulation

# Reviving Purpose Limitation and Data Minimisation in Data-Driven Systems

Asia J Biega* & Michèle Finck**

This paper determines whether the two core data protection principles of data minimisation and purpose limitation can be meaningfully implemented in data-driven systems. While contemporary data processing practices appear to stand at odds with these principles, we demonstrate that systems could technically use much less data than they currently do. This observation is a starting point for our detailed techno-legal analysis uncovering obstacles that stand in the way of meaningful implementation and compliance as well as exemplifying unexpected trade-offs which emerge where data protection law is applied in practice. Our analysis seeks to inform debates about the impact of data protection on the development of artificial intelligence in the European Union, offering practical action points for data controllers, regulators, and researchers.

## 1. Introduction

Questions around data management and analysis have been at the fore of policy debates in the European Union in recent years. A particular tension exists between the continued desire to protect personal data through a robust legal regime in the form of the General Data Protection Regulation ('GDPR'), which renders some forms of data collection and analysis unlawful, as well as the objective to generate and analyse more (personal) data so that Europe can remain competitive in the global 'data battle'.[1] There are thus simultaneous policy incentives to process both less and more personal data. This tension will accelerate in the near future with recent legislative developments including the proposed Data Governance Act and the expected Data Act.

This tension can also be pinpointed in relation to debates regarding the legal principles of purpose limitation and data minimisation. While the GDPR has affirmed both principles as core tenets of European data protection law, voices from the private sector, policy circles and academia have argued that these objectives cannot be fulfilled while reaping the benefits of "big data". Our paper examines, through an interdisciplinary law and computer science lens, whether data minimisation and purpose limitation can be meaningfully implemented in data-driven settings, in particular algorithmic profiling, personalisation and decision-making systems. Our analysis reveals that the two legal principles continue to play an important role in managing the risks of personal data processing and that they may even increase the robustness of AI systems by reducing noise in the data. These findings allow us to rebut claims that they have become obsolete.

The paper further highlights that even though these principles are important safeguards in personalisation, profiling, and decision-making systems, there are important limits to their practical implementation. Contrary to what is often claimed, these limits do not so much relate to the quantities of the processed data. Rather, we highlight that the practical difficulties of implementing data minimisation and purpose limitation are due to (A) the difficulties of measuring law and the resulting open computational research questions as well as a lack of concrete guidelines for practitioners; (B) the unacknowledged trade-offs between various GDPR principles, in particular between data minimisation and fairness; (C) the lack of practical means of removing personal data from trained models without considerable economic and environmental costs, and (D) the insufficient enforcement of data protection law.

## 2. Sources of Disagreement about Purpose Limitation and Data Minimisation

Purpose limitation and data minimisation have been proclaimed to stand in tension with data-driven business models such as those underlying profiling, personalisation and decision-making systems. Arguments against the principles range from technical infeasibility all the way to potentially causing systemic harms to the European economy. At the same time, the principles have been reaffirmed by the GDPR as they limit the collection of unnecessary data in anticipation of potential harms, and aim at maintaining a power balance between the data subjects and data controllers. This section provides an overview of those arguments and makes the case for reviving the discussion about these principles in the context of data-driven systems.

---

1  Janosch Delcker, 'Thierry Breton: European companies must be ones profiting from European data' (*Politico*, 19 January 2020) https://www.politico.eu/article/thierry-breton-european-companies-must-be-ones-profiting-from-european-data accessed 31 January 2020. Note that the new Commission now sees Europe's competitive edge in industrial rather than personal data.

*   Asia J Biega is a tenure-track faculty member and head of the Responsible Computing group, Max Planck Institute for Security and Privacy
**  Michèle Finck is Professor of Law and Artificial Intelligence, University of Tübingen. This research was carried out while she was a Senior Research Fellow at the Max Planck Institute of Innovation and Competition, where she remains an Affiliated Fellow.

## 2.1 Recent Policy Debates

Our choice to focus on algorithmic profiling, personalisation, and decision-making systems is motivated by two reasons. First, personalisation and profiling have already become key features of many online services and are likely to become even more prominent as an increasing number of online products are accompanied by a service component, a phenomenon referred to as "servitisation".[2] Second, personalisation, profiling and decision-making systems oftentimes use large quantities of data. As such, they are an especially suitable test case to examine the contemporary relevance of data minimisation and purpose limitation.

Personalisation, profiling, and decision-making systems collect personal data in the form of not only user attributes (such as gender or location), but also behavioral interaction logs (such as search queries, product ratings, browsing history, or clicks). The entirety of this data can be used to personalize search ranking results based on past clicks, to personalize product recommendations based on past product ratings, to target ads based on past visited websites, or to make decisions regarding individuals based on topical interest profiles. Thus, a variety of machine learning and data mining setups, including search, recommendation, and classification, fall within the scope of this paper.

Many current uses of machine learning ("ML") in industrial contexts are based on the repurposing of data and legal limitations thereto have been criticized. Mayer-Schönberger and Padova argued that for big data 'to reach its potential, data needs to be gathered at an unprecedented scale whenever possible, and reused for different purposes over and over again'.[3] Voss and Padova noted that 'there is one necessary condition for enabling innovation to flourish: allowing data to be processed without a pre-determined purpose'.[4] According to Moerel and Prins, 'due to social trends and technological developments (such as Big Data and the Internet of Things) the principle of purpose limitation will have to be abandoned'.[5] There are indeed scenarios where the repurposing of data has benefits, such as where speech recordings of voice-operated devices are used to train algorithms seeking to predict information about the health of the speaker.[6]

It has similarly been argued that data minimisation is no longer implementable in settings that generate value from the processing of large quantities of personal data. The incentive in the contemporary data economy is to maximise the accumulation and analysis of personal data. Some consider that with the ubiquitous generation of data the focus should lie on the use rather than the collection of data.[7] Others have pinpointed the dissonance between practices focused on the continuing accumulation of data and the legal principle. Koops has asked: '[w]ho in his right mind can look at the world out there and claim that a principle of data minimisation exists?'[8] Industry organizations have warned that Europe is 'shooting itself in the foot' with limitations on data usage in relation to AI.[9] Others consider that adhering to data minimisation 'would sacrifice considerable social benefit' as it may limit the innovative potential of ML.[10] Yet, data-driven systems present benefits as well as harms and legal intervention can address the latter.

## 2.2 Computational Evidence

Arguments of non-implementability of data minimisation in contemporary data-driven systems urge an investigation into what computational evidence has to say. On the one hand, the availability of big data has observably enabled progress in machine learning.[11] On the other, we also find evidence demonstrating feasibility of data limitation as well as algorithmic techniques that, in effect, reduce the quantity or the quality of the underlying data.

### 2.2.1 Minimising the Quantity of Data

Empirical evidence suggests that, in many data-driven settings, using increasingly larger amounts of data leads to diminishing returns in model performance. For example, in 2008, Krause and Horvitz showed that collection of additional user features leads to diminishing returns in the quality of personalized search.[12] Similar trends have since been demonstrated across a variety of ML domains, for instance, in deep learning and its applications ranging from machine translation, through language modeling, to image and speech recognition,[13] in computer vision algorithms,[14] as well as personalised recommendations.[15]

Beyond data retention heuristics focusing on performance-related properties of data, a more straightforward strategy is to retain the most recent data while discarding old data. Research-wise, the effi-

---

2   Think, for instance, of a "smart" electronic toothbrush connected to an app that offers personalised dental hygiene and toothpaste suggestions to the user.
3   Viktor Mayer-Schönberger and Yann Padova 'Regime Change? Enabling Big Data Through Europe's New Data Protection Regulation' (2016) 17 *The Columbia Science and Technology Law Review* 315, 317.
4   Axel Voss and Yann Padova, 'We need to make big data into an opportunity for Europe' (*Euractiv*, 25 June 2015) https://www.euractiv.com/section/digital/opinion/we-need-to-make-big-data-into-an-opportunity-for-europe accessed 17 January 2020.
5   Lokke Moerel and Corien Prins, 'Privacy for the Homo Digitalis: Proposal for a New Regulatory Framework for Data Protection in the Light of Big Data and the Internet of Things' (25 May 2016) 2 https://papers.ssrn.com/sol3/papers.cfm?abstract_id=2784123 accessed 17 January 2020.
6   International Working Group on Data Protection in Telecommunications, Working Paper on Privacy and Artificial Intelligence, 64th Meeting, 29-30 November 2018, Queenstown (New Zealand), 675.57.14, p 9.
7   Joris van Hoboken, 'From Collection to Use in Privacy Regulation? A Forward Looking Comparison of European and U.S. Frameworks for Personal Data Processing' in Bart van der Sloot, Dennis Broeders, Erik Schrijvers (eds), *Exploring the Boundaries of Big Data* (Amsterdam University Press 2016).
8   Bert-Jaap Koops, 'The Trouble with Data Protection Law' (2014) 4 *International Data Privacy Law* 250, 256.
9   'Artificial Intelligence: How Europe Is Shooting Itself in the Foot with the GDPR' (Fedma) https://www.fedma.org/2018/07/artificial-intelligence-how-europe-is-shooting-itself-in-the-foot-with-the-gdpr accessed 3 December 2020.
10  Mark MacCarthy, 'In Defense of Big Data Analytics' in Selinger et al (eds), *The Cambridge Handbook of Consumer Privacy* (CUP 2018) 56.
11  Alon Halevy, Peter Norvig and Fernando Pereira, 'The Unreasonable Effectiveness of Data' (2009) 24 *IEEE Intelligent Systems* 8.
12  Andreas Krause and Eric Horvitz, 'A Utility-Theoretic Approach to Privacy in Online Services' (2010) 39 *Journal of Artificial Intelligence Research* 633.
13  oel Hestness and others, 'Deep Learning Scaling Is Predictable, Empirically' [2017] arXiv:1712.00409 http://arxiv.org/abs/1712.00409 accessed 9 July 2021.
14  Chen Sun and others, 'Revisiting Unreasonable Effectiveness of Data in Deep Learning Era' [2017] *Proceedings of the IEEE International Conference on Computer Vision* (IEEE 2017) https://openaccess.thecvf.com/content_iccv_2017/html/Sun_Revisiting_Unreasonable_Effectiveness_ICCV_2017_paper.html accessed 25 July 2021.
15  Divya Shanmugam and others, 'Learning to Limit Data Collection via Scaling Laws: Data Minimization Compliance in Practice' [2021] arXiv:2107.08096 [cs] http://arxiv.org/abs/2107.08096 accessed 25 July 2021 cacy of this strategy has been demonstrated in recommender system simulations.[16] One might expect this strategy to perform well especially in settings where user behavior characteristics and preferences change over time and discarding old data might help systems keep user models up to date. One of the recent changes in Google's data retention policy, whereby web activity of new users will by default be deleted after 3 or 18 months,[17] suggests this strategy might be effective in industrial practice as well.

Despite the promise of computational feasibility, data minimisation can lead to unanticipated consequences for both the users and the service providers. Even if limiting quantities of data might lead to little accuracy loss at an aggregate level, studies have shown that data limitation would impact individual users or demographic groups to a different extent, raising the question of what data minimisation might mean in terms of fairness.[18] Minimisation of sensitive attributes has furthermore been shown to hinder the capacity of service providers to audit fairness in personalized products. Last but not least, algorithms exhibit different levels of robustness to data minimisation,[19] raising the question of how limitation obligations would impact different service providers and whether the scale of this impact would depend on how complex or state-of-the-art their algorithms are.[20]

### 2.2.2 Minimising the Quality of Data

Effects of limiting quantities of data are only one of the sources of disagreement about the desirability of data minimisation. Several studies have shown it is similarly possible to reduce the *quality* of data without reducing its overall quantity. Biega et al.'s simulations demonstrated that it might be possible to achieve good levels of personalisation for search and recommendation while randomly shuffling data (search queries or product ratings) in user profiles under certain accuracy constraints.[21] Similar techniques have been adopted to show the feasibility of such data shuffling techniques in online social communities.[22] Effectively, approaches like these allow a system to retain the volume of data and preserve system accuracy while minimising the quality of aggregated user data profiles. Other architectures have been proposed in which a user's data resides on their local device while only more crude aggregate data is shared with service providers on a need-to-know basis. The feasibility of such algorithmic architectures has been demonstrated for personalized search (a user shares only high-level topical categories describing their interests with a service provider who personalizes search results)[23] and recommendation (a local tool advises the user whether to share product click and ratings with a recommendation provider based on a privacy-utility analysis).[24] More generally, a local-device distributed learning paradigm called federated learning is an active area of research.[25]

### 2.2.3 Data-Minimising Algorithmic Techniques

A number of algorithmic techniques *de facto* minimise data. Such techniques include, for instance, outlier detection (for identifying and removing noise and rare anomalies in data), feature selection (for removing features which do not contribute or hurt the learning task), or active learning (for incrementally selecting data to be labelled or added to a model). These strategies were not developed for compliance with the legal principle of data minimisation but rather to help increase the quality of ML models or reduce data acquisition costs. Yet, they do in effect reduce the quantity of data a model uses, demonstrating that, in certain cases, data limitation might result in improved models.

Several recent papers begin to investigate how to adapt these algorithmic techniques to comply with the requirement of data minimisation. Shanmugam et al. propose a framework for automatically learning data collection stopping criteria based on an algorithm's predicted performance curve.[26] The framework adapts to different underlying feature acquisition techniques, including random as well as active learning error-reducing strategies. Goldsteen et al. leverage data anonymisation techniques to suppress and generalise input features in classification.[27] As a result, at the inference stage, a classifier has access to data of reduced quality (feature generalisation), as well as less data overall (feature suppression).

### 2.3 Benefits and Harms of Data-Driven Systems

The success and acceptance of algorithmic profiling, personalisation, and decision-making systems by both individual users and organizations that develop and deploy them speak to their benefits. Individuals may enjoy an increased quality of digital services, with personalized product recommendations, relevant ads, or search results that surface content satisfying user information needs and effectively helping sift through information overload. More effective profiling may help optimize online marketplaces and help platforms better match content consumers and producers. Profiling may also help organizations with better classification and decision-making. In certain scenarios, where classification and profiling are used to distribute a limited resource, systems may be able to allocate the resource more optimally. On a population level, behavioral data collected through search and online systems could aid developments of societally beneficial solutions for healthcare and well-being improvement, such disease outbreak predictions or detection of disease symptoms.

Personalisation, profiling and decision-making systems are subject to regulatory constraints as they can also result in a range of individual and societal harms.[28] The storage of data in those systems poses privacy risks because of potential security breaches or inadequate technical and organisational measures adopted by the data controller. For instance, if an anonymized version of the data is shared,[29] identity linking is still possible because perfect anonymisation is often infeasible.[30] Moreover, the richness of observations about a whole population of users often enables inference of additional information about individuals that is not explicitly present in the data. For this reason, it is challenging to assess the implications of data processing *ab initio*. Feasibility of inference for attributes including political convictions, sexual preferences, or personality traits has been demonstrated for social media data[31], movie rating data[32], query suggestions or targeted ads.[33] Systems might moreover make incorrect inferences about a person because of the inherent system inaccuracy or when more than one person uses the same device or account.[34] Incorrect inferences may lead to a range of consequences, from user embarrassment, to unfair denial of opportunities. Indeed, due to the various risks of personalised behavioral advertising, for instance, some have called for bans.[35]

---

16  Hongyi Wen and others, 'Exploring Recommendations under User-Controlled Data Filtering' [2018] *Proceedings of the 12th ACM Conference on Recommender Systems* (Association for Computing Machinery 2018) https://doi.org/10.1145/3240323.3240399 accessed 25 July 2021; Asia J Biega and others, 'Operationalizing the Legal Principle of Data Minimization for Personalization' [2020] *Proceedings of the 43rd International ACM SIGIR Conference on Research and Development in Information Retrieval* (Association for Computing Machinery 2020).
17  Google, 'Keeping Your Private Information Private' (*Google Blog*, 24 June 2020) https://blog.google/technology/safety-security/keeping-private-information-private accessed 3 December 2020.
18  Hongyi Wen and others (n 16); Asia J Biega and others (n 16).
19  Asia J Biega and others (n16).
20  Xavier Amatriain, 'In Machine Learning, What Is Better: More Data or Better Algorithms' https://www.kdnuggets.com/2015/06/machine-learning-more-data-better-algorithms.html accessed 8 July 2021.
21  Asia J Biega, Rishiraj Saha Roy and Gerhard Weikum, 'Privacy through Solidarity: A User-Utility-Preserving Framework to Counter Profiling', [2017] *Proceedings of the 40th International ACM SIGIR Conference on Research and Development in Information Retrieval* 675.
22  Sedigheh Eslami and others, 'Privacy of Hidden Profiles: Utility-Preserving Profile Removal in Online Forums' [2017] *Proceedings of the 2017 ACM on Conference on Information and Knowledge Management* 2063
23  Xuehua Shen, Bin Tan and ChengXiang Zhai, 'Privacy Protection in Personalized Search' (2007) 41 *ACM SIGIR Forum* 4.
24  Rachid Guerraoui, Anne-Marie Kermarrec and Mahsa Taziki, 'The Utility and Privacy Effects of a Click' [2017] *Proceedings of the 40th International ACM SIGIR Conference on Research and Development in Information Retrieval* 665.
25  Tian Li and others, 'Federated Learning: Challenges, Methods, and Future Directions' (2020) 37 *IEEE Signal Processing Magazine* 50..
26  Divya Shanmugam and others (n 15).
27  Abigail Goldsteen and others, 'Data Minimization for GDPR Compliance in Machine Learning Models' [2020] arXiv:2008.04113 [cs] http://arxiv.org/abs/2008.04113 accessed 27 July 2021.

Further harms arise because of a high complexity and a lack of transparency of data-driven systems. Many users do not understand how the systems work or what happens to their data[36], which might result in a loss of control or a sense of helplessness and powerlessness. Further feelings of unease might stem from perceptions of surveillance and a loss of privacy.[37]

On a societal level, risks of profiling and personalisation include increased surveillance, targeted censorship in authoritarian regimes, or filter bubbles.[38] Profiling can moreover reinforce 'different forms of social, cultural, religious, legal and economic segregation and discrimination' and enable the microtargeting of individuals in a manner that may profoundly affect their lives. The fact that optimisation processes inevitably prioritise certain values over others shapes online environments in a manner that can be detrimental. Beyond, ML is considered to 'influence emotions and thoughts and alter an anticipated course of action, sometimes subliminally' which may affect not only economic choices but also social and political behaviours, particularly if used without democratic oversight or control. Sub-conscious and personalised levels of algorithmic persuasion may moreover have significant effects on the cognitive autonomy of individuals and their right to form opinions and take independent decisions.[39] Recital 75 GDPR explicitly recognises that personal data processing carries risks, particularly where it serves to create 'personal profiles'.

There are moreover major risks inherent in the business model that underlies personalised advertising. Hwang has pointed out that most 'free' online services are currently financed by advertising revenue.[40] However, it is possible that personalised advertising may have little to no benefit compared to non-personalised alternatives.[41] The realisation that personalised advertising is not, in fact, superior to non-personalised alternatives such as contextualized advertising or other events such as an economic crisis may lead to a withdrawal of income for online service providers, which as a consequence no longer have a means of sustaining their operations. This would dramatically affect online services as we know them. These are more than hypothetical risks in the aftermath of a global pandemic that may trigger economic depression and result in a dramatic cut in corporate advertising budgets, particularly in light of increasing evidence that the benefits of personalised advertising hardly outweigh non-personalised alternatives.[42] Seen from this perspective, reliance on personalisation equals systemic risk.

## 3. Purpose Limitation

In essence, purpose limitation requires that the data controller define *ab initio* the purpose(s) for which personal data will be processed. This pre-defined purpose should not be exceeded, save where the new purpose is sufficiently approximate to the initial purpose or where there is an additional legal basis for further processing such as data subject consent or the need to process data for purposes such as scientific research. Purpose limitation is as old as data protection law itself and essentially serves the goal of minimising the risks that arise where personal data is processed in confining the possibilities of its usage by limiting instances of lawful processing. This section introduces the purpose of personal data processing from a legal, practical, and computer science perspective.

### 3.1 The Legal Obligation to Define a Purpose

Article 8(2) of the Charter of Fundamental Rights provides that 'data must be processed fairly for specified purposes'[43] and according to Article 5(1)(b) GDPR personal data shall be:

collected for specified, explicit and legitimate purposes and not further processed in a manner that is incompatible with those purposes; further processing for archiving purposes in the public interest, scientific or historical research purposes or statistical purposes shall, in accordance with Article 89(1), not be considered to be incompatible with the initial purposes

Article 5(1)(b) lists two distinct components: (i) purpose specification and (ii) compatible use. Purpose specification requires that personal data should only be collected for 'specified, explicit and legitimate purposes' whereas compatible use mandates that personal data shall not be 'further processed in a manner that is incompatible with those purposes.'[44] Ultimately, purpose limitation serves to manage the risk that inevitably arises when personal data is processed.

### 3.1.1 Purpose Specification

Pursuant to Article 5(1)(b) GDPR, the data controller must (i) communicate the purposes for which data is processed, which must be (ii) explicit and (iii) legitimate. This forces controllers to precisely define what data they need and discourages the accumulation of personal data for speculative future use. Importantly, the specification ought to occur before data collection (or any other processing) starts.[45]

Purpose specification can be broken down into three distinct requirements. First, the purpose must be *specified*, meaning that it must be sufficiently precise to enable the implementation of data protection safeguards and be useful to the data subject.[46] The Article 29 Working Party considers that general statements such as 'improving user experience', 'for commercial purposes' or 'for advertising' are generally not specific enough.[47] This finding is important as personalisation, profiling, and decision-making algorithms generally process personal data seeking to improve rather than simply provide a service, a description that fails the specificity test. Yet, if, as per the Working Party, improvements of user experience are not a valid purpose, it is worth wondering whether improvements in service delivery can ever be.

Such definitions are, however, contextual as the level of detail required will depend on the specific context.[48] In scientific research broader formulations are permissible as it is often not possible to fully identify the purpose at the time of data collection.[49] National supervisory authorities have in the past taken enforcement action against definitions of purpose considered to be insufficiently specific. In 2014, the Dutch DPA imposed a cease and desist order on Google, arguing that 'the provision of the Google service' was not specific enough.[50] In 2020, the Spanish supervisory authority fined a bank for violation of purpose limitation when it processed customer data for sixteen years after the end of the corresponding business relationship.[51] In 2021, the Belgian supervisory authority held that a school's parental mailing list, which did not make use of the blind carbon copy ('BCC') function, breached Article 5(1)(b) GDPR as it was not necessary to circulate all parent emails to achieve the informational purposes.[52] The literature has also drawn attention to violations of purpose specification in online advertising.[53]

Data subjects' expectations must also be accounted for. In principle, this is a laudable perspective as it takes into account the interests of the data subject. Yet, the principle's usefulness can also be questioned as data subjects' understanding of contemporary data ecosystems is extremely limited.[54] One may also wonder whether over time, the principle eradicates its own usefulness. As data collection and use practices change, so do expectations. Many current practices would likely not have been acceptable in the 1990s, whereas in the future people might be accepting of practices that would now be seen as crossing a red line. More extreme data processing may thus ultimately result in more acceptance thereof.

Second, the purpose must be *explicit*, meaning that it 'must be sufficiently unambiguous and clearly expressed'. This requires that the purposes 'must be clearly revealed, explained or expressed in some intelligible form'.[55] Where this is not the case, factual elements, common understandings and reasonable expectations are considered to determine the actual purpose.[56] Whereas purposes must be specifically defined, they should also be understandable to data subjects. To achieve both ends, layered notices are encouraged as they can both provide an overall explanation and sufficient granularity.[57] Requiring information to be explicit also underlines the connection between the purpose limitation and transparency principle, according to which data subjects must be provided with 'concise, transparent, intelligible and easily accessible' information about personal data processing.[58]

Third, the purpose ought to be *legitimate*. Legitimacy mandates that processing occurs in line with applicable law such as non-discrimination, criminal or employment law.[59] All elements of EU and national law (including municipal decrees and case law) must be respected. Legitimacy may also require respect of 'customs, codes of conduct, codes of ethics[60], contractual arrangements and the general context and facts of the case'.[61] Whereas reliance on such elements would depend on context, this is an interesting point in particular in light of the spread of AI 'ethics codes', which are designed as non-binding instruments.

The information to be provided is contextual: a small shop does not need to provide as much detail as a transnational company.[62] Where a broad user group across different cultures is targeted, information needs to be particularly clear and where a controller provides different services (such as email, social networking and photograph, video and music uploads) granularity is needed to make sure the information provided is sufficiently clear.[63] Where services are offered to particular groups such as the elderly or asylum applicants, their specific charac-

---

44   Ibid, 3-4.
45   Article 4(2) GDPR adopts a broad definition of 'processing' to include 'any operation or set of operations which is performed on personal data or on sets of personal data'.
46   GDPR, recital 39. See also Article 29 Working Party, Opinion 03/2013 on purpose limitation (WP 203) 00569/13/EN, 15-16.
47   Article 29 Working Party, Opinion 03/2013 on purpose limitation (WP 203) 00569/13/EN, 16.
48   Ibid.
49   GDPR, recital 33.
50   Joris van Hoboken (n 7).
51   https://gdprhub.eu/index.php?title=AEPD_-_PS/00076/2020.
52   https://gdprhub.eu/index.php?title=APD/GBA_-_03/2021.
53   https://hal.inria.fr/hal-02566891/document
54   Catherine Miller, Rachel Coldicutt and Hannah Kitcher (n 36).
55   Article 29 Working Party, Opinion 03/2013 on purpose limitation (WP 203) 00569/13/EN, 17.
56   Ibid, 19.
57   Ibid, 16.
58   GDPR, art 12(1) and recital 58 .
59   Article 29 Working Party, Opinion 03/2013 on purpose limitation (WP 203) 00569/13/EN, 12.
60   For a critique of favoring ethics codes over law, see http://ejlt.org/article/view/722/978#_ednref26
61   Article 29 Working Party, Opinion 03/2013 on purpose limitation (WP 203) 00569/13/EN, 20.
62   Ibid, 51.
63   Ibid, 51.

teristics need to be accounted for.[64]

It is finally worth noting that personal data can be collected for more than one purpose. Where these purposes are related, an 'overall purpose' can be used (under whose umbrella a number of separate processing operations take place), yet controllers must be careful not to identify a broad purpose in view of undertaking further processing that is only remotely related to the initial purpose.[65]

Purpose limitation is an essentially procedural requirement which does—with the exception of the legitimacy requirement—not appear to have a substantive facet. Seen from this perspective, it would mainly be an exercise in skilled legal drafting as any legitimate purpose that is formulated with sufficient specificity and explicitness would be GDPR-compliant. It is for instance worth wondering whether in the Belgian school email case referenced above, the school could have sent such an email had it defined the purpose not just as information but also networking between parents. Seen from this perspective, purpose limitation is a mindfulness exercise for data controllers in that it obliges them to ponder the use of personal data and be explicit about what it is being used for. It creates a reasonableness requirement for personal data usage in that the definitions of purpose that are too broad, including (as seen above) 'improving user experience', which is what many algorithmic personalisation, profiling and decision-making systems do, is too broad. However, if, say, a video streaming provider were to use better-skilled legal drafting to state that the purpose of processing is to 'provide personalized video recommendations', that would very likely be satisfactory. It follows that if skilled legal drafting can define the purpose in reasonable ways, it will likely pass the purpose specification test. As such, purpose specification does not genuinely limit the ways in which personal data can be used.

Both the computational and practical perspectives have revealed that personalisation and profiling usually serve to improve the service that is delivered. As per existing regulatory guidance, this is not a specific enough purpose. In order for such practices to become aligned with data protection law, they need to be more *specific* which can be achieved by more detailed, or layered statements. Yet, if more precise language is all that is needed to ensure compliance with purpose specification, it is worth wondering what objective it actually serves in data protection law. Can any purpose be used (as long as legitimate and explicit) provided that it is put in precise language? If so, what objective does purpose specification actually fulfil? This would indicate that any purpose can be realised as long as it is formulated specifically and corresponds to other data protection requirements.

### 3.1.2 Compatible Use

The second component of purpose limitation is compatible use. It requires that personal data be not further processed in a manner incompatible with the original purpose(s). However, the mere fact that data is processed for a purpose different from that originally defined does not mean that it is automatically incompatible.[66] In some circumstances, processing for a different purpose is considered sufficiently connected to the original purpose. This requires a case-by-case evaluation of whether the initial and further processing are compatible. Here, relevant criteria according to the Article 29 Working Party's 2013 guidance are (i) the relationship between the different purposes; (ii) the context of collection and the reasonable expectations of data subjects; (iii) the nature of personal data and the impact of further processing on data subjects; and (iv) the safeguards adopted by the controller.[67] National DPAs have also issued guidance on the interpretation of compatible use. The UK Information Commissioner's Office ('ICO') for instance considers that the new use must be 'fair, lawful and transparent'.[68]

To assess whether further processing was implied in the original purpose, adopting the perspective of 'a reasonable person in the data subject's position' has been recommended.[69] However, as observed above, consumers rarely have a realistic understanding of contemporary processing practices. Moreover, the nature of the contract and the relation between the data subject and the data controller are to be accounted for.[70] For example, the public disclosure of personal data is a relevant factor.[71]

Processing that is incompatible with the original purpose cannot be legitimized through reliance on an alternative legal ground under Article 6 GDPR.[72] In principle, data controllers would thus have to anonymise personal data to process it beyond the limited purpose (as this would bring the data outside of the scope of the GDPR).[73] This is, however, often easier said than done considering the difficulties of achieving anonymisation.[74] The compatible use requirement is a much stronger practical constraint on data processing than purpose specification, which, as seen above, is largely an exercise in skilled legal drafting. Once the purpose has been specified, however, compatible use does impose considerable practical constraints on the possibilities of data use. This is a general challenge in respect of the European Commission's current policy agenda that seeks to further incentivise the sharing of data.[75] In our context, it for instance applies that where the purpose of the collection of an address is specified as necessary for billing purposes, this information cannot be used to inform personal recommendations on the basis of location. There are, however, a number of additional instances where data can be processed beyond the purpose.

A question remains of when two purposes are compatible. When the processing purpose is stated to be an improvement in performance and the performance is measured using well-defined metrics, one natural computational interpretation of compatible use is when metrics are positively correlated. Consider the following example. An online outdoors store originally collected personal data such as product ratings to improve the performance of personalized hiking gear recommendations. The store expands its catalogue to include hiking clothing, and it turns out that shopper preferences for certain categories of clothing and certain categories of gear are correlated (e.g., producer brand, price range, or other features used for person-alisation). The data lawfully collected for the purpose of improving personalized gear recommendation will likely improve the clothing recommendations, thus it could be argued that it can be used for the latter purpose under compatible use. Practical implementations would need to account for various challenges, including the existence of spurious correlations or handling of implicit latent features in the algorithms.

---

64　Ibid.
65　Ibid, 16.
66　Ibid, 21.
67　Ibid, 23-26.
68　Information Commissioner's Office, 'Guide to the General Data Protection Regulation (GDPR) – Principle (b): Purpose limitation' https://ico.org.uk/for-organisations/guide-to-data-protection/guide-to-the-general-data-protection-regulation-gdpr/principles/purpose-limitation accessed 10 January 2020.
69　Article 29 Working Party, Opinion 03/2013 on purpose limitation (WP 203) 00569/13/EN, 24.
70　Ibid.
71　Ibid, 26.
72　Article 29 Working Party, Opinion 03/2013 on purpose limitation (WP 203) 00569/13/EN, 3.
73　Ibid, 7.
74　Luc Rocher, Julien Hendrickx and Yves-Alexandre de Montjoye, 'Estimating the success of re-identifications in incomplete datasets using generative models' (2019) 10 *Nature Communications* 3069 and Michèle Finck and Frank Pallas (n 30).
75　Such as through the proposed Data Governance Act.

## 3.2  Current State

To understand the purpose limitation *status quo*, we consider how key service providers using ML define the purpose of personal data processing. Google has a lengthy layered description. It first informs users that data is collected 'to build better services' before providing examples of what this means.[76] In relation to personalisation, information is used to provide services such as 'recommendations, personalized content, and customized search results' as well as personalized ads (depending on a user's settings). Google further informs its users that it uses 'automated systems that analyze your content' to provide customized search results or ads. Facebook also adheres to a layered approach by informing users that personal data is used to 'provide and support the Facebook Products', which includes 'to personalise features and content (...) and make suggestions for you (...) on and off our Products'.[77] Netflix informs its users that it processes personal data to: (i) receive newsletters, (ii) send push notifications, (iii) enhance customer experience, and (iv) fulfil legal or contractual obligations.[78] These are but some examples that highlight that the purpose of data processing can be defined in a number of ways, in line with the service provider's product and objective. As will be seen below, a weakness of the current system is that, despite publically available data processing policies and development of automated tools that can analyze them,[79] there are virtually no means of checking whether these verbal expressions correspond to what happens in practice, highlighting the difficulties of practically enforcing data protection law.

## 3.3  Service Improvement as Purpose Specification

Service providers state that user interaction data is collected to provide, improve, or personalize their services. Yet, some of these purposes are in fact not well-grounded in computational practice, as it is not immediately clear whether and which data is actually necessary to improve service results.

Firstly, it is crucial to observe that, from a system's perspective, ongoing collection of user data is not necessary to *provide* services such as search, recommendation, or classification. In web search, a ranking of webpages can be computed by matching keywords in queries to words appearing in web pages. In fact, ranking methods based on properly weighted word statistics beat some of the more complex methods in search benchmarks.[80]  Moreover, while processing a query might be necessary to complete a given search transaction, it might not be necessary to store it for future use once the search task is complete. In their search personalisation audit studies, Hannak et al. have not observed search-history-based personalisation in Google or Bing.[81] Recommendations can be unpersonalized and based on external data (for instance, sales statistics of cinema tickets could be used as a popularity indicator for recommending movies) or even random. Thus, *providing* a service in such scenarios does not appear to be a valid purpose for collection of user data. Instead, user data in personalisation, profiling and decision-making systems is collected to *improve* the results. Service improvement could be considered an objective criterion for purpose formulation if it were legitimate, explicit and specific enough.

Determining whether improvement of a service is legitimate could be thought of as conditional on the legitimacy of the service itself. In most cases, improving a legitimate service might be considered legitimate as well.

Explicitness, as argued by Koops,[82] could be enhanced by specifying purposes in a machine-readable format such as XML. Indeed, such code-driven expression forces data controllers to reflect on their processing goals explicitly. Fouad et al. has proposed to improve explicitness of browsing cookie purposes by listing them in a structured table in the data processing policy.[83] This approach thus allows for quick identification of processing purposes.  A solution along these lines appears viable for data collection purposes in data-driven systems as well.

As for specificity, however, our earlier analysis revealed the question of whether service improvement can be considered a purpose specific enough (since, as per the Working Party, improvement of user experience is not).[84] We thus consider ways in which improvement can be stated more concretely to pass the specificity test.

Von Grafenstein has argued that purpose standardisation aids in increasing purpose specificity and legal certainty, as 'both the individuals concerned and data controllers, which are part of this "purpose"-oriented system, are reassured that all data processing occurs under the same conditions.'[85] To this end, Fouad et al. proposed using ontologies not only to standardize purpose descriptions but also to allow reasoning about ontological relations between purposes, such as subsumption.[86]

In the context of data-driven systems, relations between improvement purposes could be defined along the axes of *what* and *how*. Increasing specificity along the *what* axis might entail defining which functionality in the system would exactly be improved. For instance, a layered description might indicate improvements in personalized search, and within that purpose, specify which topics of search queries will see improvement in the results. Increasing specificity along the *how*

---

axis might entail specifying how improvement will be quantified. For instance, the goal of a system might be to display relevant documents a certain number of ranking positions higher in the search results than they would be without the collected personal data. Neither of the suggested directions are straightforward to implement, however.

### 3.4 Computational Challenges

*Improvement* is an ambiguous concept and needs additional specification in practical and computational terms. Generally, it is reasonable to assume improvements would be quantified as *differences* in selected *system performance metrics*. Performance evaluation is widely used to judge models and systems in scientific publications, to measure progress in the field through public benchmarks[87], or to determine if updates to tech products should be shipped using techniques such as A\B testing[88]. A natural consequence would thus be to similarly reason about purpose limitation via performance and *tie the purpose of data collection to improvements in system performance*. Practically, to form a quantitative basis of purpose, it remains to be determined (i) which metrics to choose, (ii) how to obtain their values, and (iii) which level to aggregate metric differences at.

A metric would have to be selected from a suite of metrics that often guide system quality measurement. The main reason for such complex evaluation setups is that different metrics capture different aspects of performance, and often none of these aspects is more important than another. Moreover, individual metrics in a suite will often disagree as to whether a change leads to service improvement. Further complications include the fact that metrics might differ by application domain. A system performance will be measured differently for personalized movie recommendations than for personalized search. Finally, it is important to acknowledge that metrics often serve as simpler, quantifiable proxies for measurement targets. For instance, in systems such as search and recommendation, the goal might be to improve 'user satisfaction'. User satisfaction is, however, approximated using simpler measurable concepts such as the number of clicks.

Despite the fact that metrics are imperfect approximations of hard-to-model concepts such as 'user satisfaction', using them as a ground for purpose limitation would enable proxy metrics to determine which data should and should not be collected. Barocas and Selbst discuss the caveats behind a related concept of target variables in machine learning models deployed in societally sensitive applications.[89]

Values of performance metrics can be obtained using quantitative, qualitative or mixed evaluation methods. For instance, in personalized search the goal of a system might be to reduce the time necessary to find the desired information for ambiguous queries. Whether the system achieves this goal might be measured quantitatively using the rate of query rephrasing (when a user rephrases their query, the preceding query has likely not yielded satisfying results), or qualitatively, through an in-person user experience interview.

Finally, there might be multiple approaches to aggregating metric improvements. Among many possible options, a formal definition might require that the collection of data improves the service on average for all users, or that the collection of data improves the service for the individual from whom the data is collected.[90]

More detailed guidance will be necessary for practitioners to navigate all the discussed design choices as they are likely to lead to different minimisation outcomes.

## 4. Repurposing Personal Data Beyond Compatible Use

The purpose limitation principle requires that before any personal data processing can take place the purpose thereof be defined. Subsequent processing is assessed against that purpose. This stands in contrast with the reality of much contemporary data mining practice where the data that is mined was often initially collected for another purpose. Beyond the compatible use requirement examined above, the GDPR acknowledges other avenues of processing data beyond the initial purpose. First, the scientific research exemption recognises that in this specific context, it is often difficult to foresee potential future uses of personal data.[91] Second, the GDPR acknowledges that where personal data is further processed for 'statistical purposes', this shall not be considered incompatible with the original purpose.[92] Third, and more controversially, data controllers are also able to move beyond purpose limitation in getting data subjects to consent to further processing.

### 4.1 Scientific Research

Article 5(1)(b) GDPR foresees that personal data can also be further processed for 'scientific' purposes in accordance with the safeguards listed in Article 89 GDPR, including technical and organizational measures and respect for data minimisation. In such circumstances, member state law may also provide for derogations from some data subject rights.[93] The scientific purpose exemption shall be 'interpreted in a broad manner including for example technological development and demonstration, fundamental research, applied research and privately funded research'.[94]

Given the broad definition of scientific research it is pertinent to wonder what can be considered to be 'scientific research' in the context of personalisation, profiling, and decision-making systems. While publicly funded and externally published work at academic institutions might rather uncontroversially be considered as research, what constitutes research at private technology companies is less clear. Industrial research teams might do both internally and externally facing work, with the internal research not meant for external publication but rather for proprietary product development and innovation. To complicate matters further, various company organizational structures often make it impossible to distinguish who works on research and who works on products, despite official employee titles that might suggest a clear distinction. For example, at Google, research scientists are embedded in engineering teams[95] and as a result many research scientists develop products, and many software engineers do externally-facing research. Spotify does not follow traditional organizational hierarchies, and employees with different backgrounds and titles are organized according to various functional dimensions, for instance, system feature areas.[96]

---

87  Benchmarks are at the center of some Computer Science conferences such as TREC https://trec.nist.gov/overview.html (accessed 24 February 2020), or TAC KBP https://tac.nist.gov/about/index.html (accessed 24 February 2020), both organized by the National Institute of Standards and Technology.
88  Ron Kohavi and Roger Longbotham, 'Online Controlled Experiments and A/B Testing' in Claude Sammut and Geoffrey I Webb (eds), *Encyclopedia of Machine Learning and Data Mining* (Springer US 2017) http://link.springer.com/10.1007/978-1-4899-7687-1_891 accessed 3 December 2020.
89  Solon Barocas and Andrew D Selbst, 'Big Data's Disparate Impact' (2016) 104 *California Law Review* 671.
90  Asia J Biega and others (n 16).
91  GDPR, arts 5(1)(b) and 89.
92  See also GDPR, recital 50: 'processing for archiving purposes in the public interest, scientific or historical research purposes or statistical purposes should be considered to be compatible lawful processing operations'.
93  Article 89(2) GDPR.
94  Recital 159 GDPR.
95  Alfred Spector, Peter Norvig and Slav Petrov, 'Google's Hybrid Approach to Research' (2012) 55 *Communications of the ACM* 34.

## 4.2 Statistical Purposes

The GDPR moreover permits further processing for another purpose where that purpose is 'statistics.'[97] If an activity qualifies as statistical analysis, controllers benefit from a more favorable regime, including that data can be kept for longer than necessary for the purposes of processing.[98] The recognition of a more favorable regime for statistical analysis reflects that traditionally, data used for statistics was usually initially collected for another purpose. For example, national statistics offices have relied on data collected for other ends to carry out their work. Given the overlaps between statistics and computational learning, it is worth enquiring whether ML can be qualified as statistical analysis under the GDPR to benefit from the corresponding legal regime. Indeed, just as statistics, data used to train ML systems is often repurposed.

Recital 162 GDPR defines statistical purposes as a form of processing 'necessary for statistical surveys or for the production of statistical results.' These results may be further used for different purposes. The recital, however, also makes clear that the output must not be 'personal data, but aggregate data' and that, moreover, the results 'are not used in support of measures or decisions regarding any particular natural person'.[99] The GDPR is thus clear that some ML outputs cannot be qualified as 'statistics', namely those that generate personal data or are used to support individual measures and decisions.[100] It seems uncontroversial that personalised services are indeed an individual measure.

From the above it would seem that some forms of ML output could qualify as statistics whereas others cannot. Concretely, whereas the prediction of overall customer churn would be statistics, the prediction of whether a given customer will leave and the initiation of corresponding action (such as more attractive personalised deals) could not fall within the scope of this more favorable regime. The mere fact that statistical methods are used in private commercial settings (as opposed to statistical analysis in the public interest) nonetheless does not form a bar to the application of the statistical exemption, which does apply to 'analytical tools of websites or big data applications aimed at market research'.[101] The GDPR furthermore enables the EU or Member States to create specialised regimes on processing personal data for statistics.[102] If one or several Member States would choose this route (such as to attract data analysis companies to their jurisdiction) there is a clear risk of fragmentation in the Digital Single Market - going counter the GDPR's harmonising objective.

Although computing experts disagree whether machine learning is different from statistics, many do point out that they are indeed different to a certain extent. Some argue that they have different goals (prediction vs. inference and analysis of relations between random variables) while sharing some of the algorithms and practices.[103] [104] Others argue that the disciplines are complementary although increasingly converging.[105]

There is another distinction relevant in the context of the above discussion on individual vs. aggregate outputs. Namely, machine learning pipelines produce aggregate and individual results at different processing stages. Many of the systems considered in this article leverage large sets of user data to construct a model, and then use such an aggregate model in conjunction with an individual's data to compute the individual's results. For instance, a search engine might train a non-personalized ranker that preselects webpages as a response to a query, and then use an individual's personal data to re-rank the webpages in the preselected set. In a scenario like this, the training of the aggregate model might be considered a form of statistical analysis (and thus not subject to data minimisation), while applying the model in conjunction with an individual's data will not (as it produces individual results).

One caveat to consider is that there exist personalisation algorithms that do not conform to the above scheme. For instance, in personalized recommendation models based on matrix factorisation techniques, all individuals' data is used to train the model and no new data is used at the application stage. In such a case, a given subject's data is used both to train an aggregate model as well as to produce individual results.

## 4.3 Consent as the Silver Bullet?

Where processing goes beyond compatible use, it can be legitimized by the data subject's consent (or where it is based on EU or Member State law).[106] Subject to consent, 'the controller should be allowed to further process the personal data irrespective of the compatibility of the purposes'.[107] Today, many data controllers use consent to legitimise data processing for purposes that would otherwise not be lawful as they exceed the initial purpose.[108] Where ML is used to inform measures or decisions in relation to individuals, consent 'would almost always be required', in particular for direct marketing, behavioural or location-based advertisement, data-brokering, or tracking-based digital market research.[109] There accordingly appears to be an assumption that these types of analysis are too different from the original purpose to be legitimised by compatible use.

From this perspective, consent appears as the silver bullet to get around legal limitations of purpose limitation. However, using consent to legitimise otherwise illegitimate data processing has been

---

criticised.[110] Consumer rights organizations have pointed out that it allows data controllers to circumvent purpose limitation and makes it hard for consumers to understand contemporary data flows.[111] In general, using consent as a lawful basis for personal data processing is controversial. Consent is an expression of the paradigm of informational self-determination, designed to give data subjects 'control' over their personal data.[112] However, as underlined above, there is now broad empirical evidence questioning whether data subjects really are in a position to make such informed choices as they by and large do not understand the complexity of contemporary data flows.[113] Indeed, many individuals seem unaware of all the kinds of data processed by controllers, including what has been termed as 'bastard data': where the merging and comparing of data results in additional personal data.[114]

The challenge of acquiring informed consent for service improvements, specifically, lies in explaining the value of improved results vis-à-vis the various costs of collecting different pieces of user data. On the one hand, it is hard to expect the service provider to ask for such fine-grained consent when studies show that users have a limited understanding of the overall digital ecosystem, with some users not even aware that data such as search queries is stored and collected in the first place.[115] On the other hand, existing studies on related problems show that it is feasible to directly ask users for their privacy preferences when it comes to feature collection[116], or indirectly estimate how much users value their data in the context of specific tasks such as disease predictions.[117] At the same time, several lines of research, including explainable AI, or uncertainty and risk communication[118], aim at communicating the outputs of computational systems to end users in an understandable way. While those lines of work (on privacy preferences and outcome communication) are largely separate, it is feasible to imagine combining both approaches to design informed consent solutions for service improvements in personalisation, profiling, and decision-making systems.

Scholarship has long warned that consent 'should not bear, and should never have borne, the entire burden of protecting privacy'.[119] Consent is considered to be mainly theoretical and devoid of practical meaning, particularly since many Internet-based services cannot be used without consent.[120] It has indeed been suggested that the main feature of consent is 'to performatively legitimate otherwise unregulated unacceptable corporate practices.'[121]

There is thus broad scepticism regarding the suitability of consent as a legitimising basis. Furthermore, there is also reason to wonder whether the detailed requirements for valid consent can be met in the specific context of personalisation, profiling, and decision-making systems. The GDPR defines consent as 'any freely given, specific, informed and unambiguous indication of the data subject's wishes by which he or she, by a statement or by a clear affirmative action, signifies agreement to the processing of personal data relating to him or her'.[122] As a consequence, some forms of 'consent' such as pre-ticked boxes do not meet the GDPR threshold.[123]

In 2019, the French supervisory authority CNIL imposed a fine of €50 million on Google for having failed to get valid consent.[124] Important information such as the definition of the purpose was excessively disseminated across several pages, meaning that consent could neither be informed nor unambiguous or specific - the latter because users had to agree to the bulk of Google's terms before being able to use its services.[125] Consent appears to not be informed in most cases as a majority of users report not reading privacy policies of the services they use.[126]

Meeting the Regulation's requirements for valid consent is indeed extremely difficult in complex online data ecosystems. This can be seen in relation to real-time bidding, the process by which websites auction off personalised advertising space on websites in real-time.[127] The Interactive Advertising Bureau, a key industry organization, itself recognised that in real-time bidding, consent cannot be achieved as data subjects lack relevant information about data controllers.[128] Due to the complexity of such systems, data subjects are not in a position to understand the implications of clicking 'I agree'. In fact, research conducted in the United Kingdom in 2019 revealed that whereas 63% accept that online services are funded by advertisements, acceptance rates shift radically to only 36% once it is explained that personal data beyond browsing history is used to personalise ads.[129] This finding

---

110 Note that Article 8(2) of the Charter of Fundamental Rights refers to consent, and Treaty change would thus be necessary to remove consent as a valid ground for processing personal data.
111 Verbraucherzentrale Bundesverband, 'Zweckänderung in der EU-Datenschutzverordnung: Stellungnahme des Verbraucherzentrale Bundesverbands zum Expertengespräch zur Regelung der zweckändernden Weiterverarbeitung personenbezogener Daten in der EU-Datenschutz-Grundverordnung' (17 December 2014) https://www.vzbv.de/sites/default/files/downloads/EU-Datenschutzverordnung-BMI-Zweckaenderung-Stellungnahme-2014-12-17.pdf accessed 13 December 2019.
112 The European Commission has often underlined the GDPR's role in providing data subjects with control over personal data: European Commission, 'EU data protection rules' https://ec.europa.eu/info/priorities/justice-and-fundamental-rights/data-protection/2018-reform-eu-data-protection-rules/eu-data-protection-rules_en accessed 31 January 2020.
113 Omri Ben-Shahar and Carl Schneider, 'The Failure of Mandated Disclosure' (2011) 159 *University of Pennsylvania Law Review* 647.
114 'ENDitorial: Is "Privacy" Still Relevant in a World of Bastard Data?' (European Digital Rights (EDRi)) https://edri.org/our-work/enditorial-is-privacy-still-relevant-in-a-world-of-bastard-data accessed 3 December 2020
115 Catherine Miller, Rachel Coldicutt and Hannah Kitcher (n 36).
116 Andreas Krause and Eric Horvitz (n 12).
117 Gilie Gefen and others, 'Privacy, Altruism, and Experience: Estimating the Perceived Value of Internet Data for Medical Uses' [2020] *Companion Proceedings of the Web Conference 2020* 552.
118 David Spiegelhalter, 'Risk and Uncertainty Communication' (2017) 4 *Annual Review of Statistics and Its Application* 31.
119 Solon Barocas and Helen Nissenbaum, 'Big data's end run around procedural privacy protections', (2014) 57 *Communications of the ACM* 31, 33.
120 Bert-Jaap Koops (n 8) 251-252. , 'The Trouble with Data Protection Law' (2014) 4 International Data Privacy Law 250, 251-252.
121 Elettra Bietti, 'Consent as a Free Pass: Platform Power and the Limits of the Informational Turn' (2019) Pace Law Review (forthcoming).
122 GDPR, art 4(11) .
123 GDPR, recital 32 and Case C-673/14 Planet 49 [2019] ECLI:EU:C:2019:801.
124 Commission nationale de l'informatique et des libertés (CNIL), 'Délibération de la formation restreinte n° SAN – 2019-001 prononçant une sanction pécuniaire à l'encontre de la société GOOGLE LLC' (21 January 2019), SAN-2019-001.
125 It is worth noting that Google has appealed this decision.
126 Brooke Auxier and others, 'Americans and Privacy: Concerned, Confused and Feeling Lack of Control Over Their Personal Information' (Pew Research Center: Internet, Science & Tech, 15 November 2019) https://www.pewresearch.org/internet/2019/11/15/americans-and-privacy-concerned-confused-and-feeling-lack-of-control-over-their-personal-information accessed 3 December 2020.
127 For an overview, see Information Commissioner's Office, 'Update report into adtech and real time bidding' (20 June 2019) https://ico.org.uk/media/about-the-ico/documents/2615156/adtech-real-time-bidding-report-201906.pdf accessed 17 January 2019.
128 Johnny Ryan, 'New evidence to regulators: IAB documents reveal that it knew that real-time bidding would be "incompatible with consent under GDPR"' (*Brave*, 20 February 2019) https://brave.com/update-on-gdpr-complaint-rtb-ad-auctions accessed 17 January 2020.
129 Information Commissioner's Office, 'AdtechMarket Research Report' (March 2019) 5, 19 https://ico.org.uk/media/about-the-ico/documents/2614568/ico-ofcom-adtech-research-20190320.pdf accessed 17 indicates that, if consent really were informed, most users would not consent.

The requirement that consent be given 'freely' might also have far-reaching implications in personalisation. One may in fact wonder whether there is free consent in the absence of a non-personalised alternative. Recital 42 GDPR provides that there is no free consent 'if the data subject has no genuine or free choice or is unable to refuse or withdraw consent without detriment'.[130] Where a website cannot be used without consenting to personal data processing, 'the user does not have a real choice, thus the consent is not freely given'.[131] In 2013, the CJEU held that consent cannot be used as a lawful basis for fingerprinting in the process of obtaining a biometric passport as people need a passport and there is no alternative option available.[132] Although passports are arguably more essential than the use of specific online services, practical requirements to use the latter should not be underestimated (think, for instance, of the importance of search engines for contemporary lives or the significance of cloud computing providers for most businesses) and the Court's reasoning could also hold in relation to the latter. There thus appears to be a presumption that consent is invalid unless there is an alternative to use the service in a non-personalised way. As a consequence, consent 'should not generally be a precondition of signing up to a service'.[133] In 2018, an NGO brought a case (still pending) in Austrian courts that enquires whether consent is really free where users have no choice but to consent to continue using a service.[134]

What is more, pursuant to Article 7(4) GDPR, for consent to be freely given, 'utmost account shall be taken of whether, inter alia, the performance of a contract, including the provision of a service, is conditional on consent to the processing of personal data that is not necessary for the performance of that contract'. This indicates that there is a higher threshold for consent where it is used to justify a purpose that cannot be included in the initial purpose - itself governed by contract under Article 6(1)(b) GDPR.

For consent to be informed and to ensure transparency, 'data subjects/consumers should be given access to their 'profiles', as well as to the logic of the decision-making (algorithm) that led to the development of the profile'.[135] The requirement that data controllers disclose their 'decisional criteria' is considered particularly important as inferences can be more sensitive than the original data itself (a point we examine separately below).[136] This is an interesting statement as it may require a disclosure of the algorithm - contrary to what is generally considered necessary under Article 22 GDPR.

The feasibility of automatically checking for compliance with data processing declarations is another dimension pertinent to establishing whether consent mechanisms are meaningful. In the context of cookie processing, Santos et al. have argued for standardisation of consent in terms of both interfaces and language, as well as consent storage and withdrawal mechanisms.[137] Standardisation in this scenario might facilitate automated audits of data processing policies, enhance processing transparency, and increase the likelihood of consent being informed. The authors note, however, that in practice validating whether the consent policies are complied with will often require extensive manual validation. In data-driven systems, even if consent messaging as well as protocols were to be standardized, auditing for compliance would also require manual efforts. Crucially, these manual validations would have to be conducted in close cooperation with service providers. To the best of our knowledge, appropriate black-box auditing methods for compliance with service improvement purposes in data-driven systems have thus far not been developed.

A further practical question regarding consent as a legitimation of personalisation relates to Article 7(3) GDPR, which provides that data subjects can withdraw consent at any time. Whereas the withdrawal of consent does not negate the legitimacy of processing before withdrawal, it bars data controllers from continuing to process the data once the right has been revoked. This requirement would imply that, should a withdrawing user's data form a part of a trained model, the model might no longer be processed after consent is withdrawn.

It is far from established how to operationalise Article 7(3) GDPR in ML. Computers scientists have only recently started to develop solutions for efficient deletion of individual data points from trained machine learning models[138] and further research is necessary. At present, the complete removal of a user's data can often only be achieved by retraining the model from scratch on the remaining data, a procedure which is computationally costly and thus neither economical, practical or environmentally desirable.[139] It is worth noting that the same problem emerges where consent is exhausted once the purpose has been achieved.

Our analysis in this section has shown that the GDPR frames consent as a tool to get around purpose limitation requirements. Where an individual consents to expanded data processing, such processing can take place. This framing is problematic for a number of reasons. First, it minimises the effectiveness of purpose limitation. Second, it contributes to the increasing opacity of personal data processing (as examining purposes in terms of use rarely provides a transparent picture of what personal data is used for). Third, the specific legal requirements around consent—that it be freely given, specific, informed and unambiguous—can rarely if ever be meaningfully complied with. Yet, to date there has been insufficient enforcement of the legality of consent, as with the GDPR overall. Finally, there are currently no technical tools to efficiently implement the logical consequences of consent revocation.

---

January 2020.
130　GDPR, recital 42.
131　Eleni Kosta, 'Peeking into the cookie jar: the European approach towards the regulation of cookies' (2013) 21 *International Journal of Law and Information Technology* 380, 396.
132　Case C-291/12 Schwartz [2013] EU:C:2013:670, para 32.
133　See further Information Commissioner's Office, 'Guide to the General Data Protection Regulation (GDPR) – Consent' https://ico.org.uk/for-organisations/guide-to-data-protection/guide-to-the-general-data-protection-regulation-gdpr/lawful-basis-for-processing/consent accessed 18 October 2019; Article 29 Working Party, Guidelines on consent under Regulation 2016/679 (WP259 rev.01) 17/EN, 6; Frederik Borgesius et al, 'Tracking Walls, Take-It-Or-Leave-It Choices, the GDPR, and the ePrivacy Regulation' (2018) 3 *European Data Protection Law Review* 353, 361 (making this argument in relation to consent for tracking walls on websites).
134　Noyb, 'GDPR: noyb.eu filed four complaints over "forced consent" against Google, Instagram, WhatsApp and Facebook' (25 May 2018) https://noyb.eu/wp-content/uploads/2018/05/pa_forcedconsent_en.pdf accessed 17 January 2020.
135　Article 29 Working Party, Opinion 03/2013 on purpose limitation (WP 203) 00569/13/EN, 46.
136　Ibid, 47.
137　Cristiana Santos, Nataliia Bielova and Célestin Matte, 'Are Cookie Banners Indeed Compliant with the Law?' (2020) 2020 *Technology and Regulation* 91 https://techreg.org/index.php/techreg/article/view/43/25.
138　Antonio Ginart and others, 'Making AI Forget You: Data Deletion in Machine Learning' (2019) 32 *Advances in Neural Information Processing Systems* 3518.
139　Ibid.

### 4.4 Trade-Offs Inherent to Purpose Limitation

Purpose limitation comes with a number of considerable trade-offs. First, there is the explicit and significant trade-off between an unlimited and limited processing of personal data. The GDPR is a recent legislative affirmation of purpose limitation as a core tenet of data protection law. Data protection law ultimately serves to manage the risks that inevitably arise when personal data is processed and purpose limitation seeks to reduce such risks by limiting the ways in which the data can be processed. It has already been seen above that this limitation of data processing has in recent years been criticised as potentially stifling an innovative EU data economy, including in respect of artificial intelligence. It can be assumed that discussions about the desirability of purpose limitation will be revived in the coming years in light of envisaged legal reform (in the form of the proposed Data Governance Act and the expected AI Act) that would incentivise increased sharing and thus also repurposing of (personal) data. The promotion of data sharing services (also referred to as 'data marketplaces', essentially intermediaries that match data providers and data users) questions the very validity of purpose limitation. As such, we can expect an explicit and heated debate as to whether purpose limitation stands in the way of data sharing and the related expected societal benefits (such as in healthcare or climate change mitigation) in the EU.

Beyond this overarching explicit trade-off, our analysis has also revealed other trade-offs that were probably not envisaged by the legislative process. First, there is a trade-off between honesty and flexibility in purpose specification. It was observed that the purpose needs to be defined *ex ante*, yet any legitimate, sufficiently precise and explicit purpose meets the specification test. Data controllers might make a calculated decision as to whether to honestly define their present purpose or list different purposes not necessarily pursued in the present to cover potential future uses. Second, depending on the interpretation given to the research exemption, companies might have to make trade-offs in their organizational structures. If the exemption only applies to separate research teams, purpose limitation might disincentivise the creation of more integrated teams, even though such teams might be more beneficial in other respects.

### 4.5 Interim Conclusion

Our examination of the application of purpose limitation to personalisation, profiling and decision-making systems has revealed that purpose specification is a largely procedural criterion that does not really limit the ways in which personal data can be processed. While the compatible use requirement does aim at substantially limiting processing, there remains considerable uncertainty regarding the interpretation of the exemptions related to scientific research and statistics in data-driven systems. Furthermore, the limitations around data subject consent are not enforced in practice.

This does not, however, mean that the purpose limitation principle fulfils no function in data protection law. First, it forces controllers to ponder the need for and implications of personal data processing from the beginning. Second, respecting related requirements provides assurance to good-faith data controllers that processing is lawful. This echoes some elemental features of the GDPR, such as its role as a risk-management framework[140] (there is a recognition that processing generates risks and thus ought to be limited to what is necessary) and the balancing of the rights and interests of data subjects and controllers (in this case recognising that data controllers have an interest in processing data but their interests must be balanced against those of the data subject).[141] Nonetheless, our analysis above has shown that purpose limitation does not by itself stand in the way of profiling or personalisation systems. It does, however, result in numerous trade-offs, some of which might have been unintended. Below, we examine whether the same conclusion holds in relation to data minimisation.

### 5. Data Minimisation

Data minimisation is the logical consequence of purpose limitation. Article 5(1)(c) GDPR provides that data shall be 'adequate, relevant and limited to what is necessary in relation to the purposes for which they are processed'. It requires that no more personal data than necessary to achieve the purpose is processed and is also one of the 'technical and organizational measures' under Article 25(2), which reiterates that controllers only process personal data 'necessary for each specific purpose of the processing'. Thus, data minimisation should be engineered relative to the purposes.[142] Like purpose limitation, data minimisation is a risk-management measure as processing of excess data creates unnecessary risks from 'hacking to unreliable inferences resulting in incorrect, wrongful, and potentially dangerous decisions.'[143] Such risks can be minimised by making sure that controllers do not have more data than necessary and process it for no longer than necessary. Minimising the amount of data may even, depending on context, improve the quality of ML as there is less need to clean the data and less risk of inaccuracy (where the right data is chosen). Indeed, the quality of the training data and the features can be more determinative of model accuracy than the quantity of the training data.[144] To provide further context to these discussions, this section examines, from a legal and computational perspective, the three distinct components of data minimisation, namely that data must be (i) adequate, (ii) relevant, and (iii) limited to what is necessary in relation to the purposes for which they are processed.

### 5.1 Relevance

The GDPR requires that data processed for a given purpose be 'relevant'. Whereas this term has not been authoritatively defined, it appears to require that only pertinent data is processed.[145] Thus, a controller that processes irrelevant data breaches the principle. Imagine, for example, the scenario of an e-commerce website that requests your complete date of birth to provide personalised recommendations for future purchases. Unless its recommendations are supposed to have an astrological flavour, this data is irrelevant as indeed, it is likely that the company would be collecting this data to ends different from the stated purpose.

Seen from this perspective, relevance is designed to safeguard against the accumulation of data for the sake of gathering data or for undisclosed ends. There is no doubt that personal data has become an extremely valuable commercial asset and there are incentives for controllers to accumulate a maximum thereof to develop their own business model, for speculative later use, or to re-sell.[146] It is worth noting however that, as stated by the International Working Group on Data Protection in Telecommunications, the capabilities that AI systems provide 'are pushing the limits for what is relevant, and the push to provide more and more data to facilitate connections pushes the data minimisation principle' as data becomes more meaningful when combined with 'other data, greater processing capacity and deeper analyses'.[147] Given the risks associated with an uncontrolled accumulation of personal data, the GDPR imposes limits on such practices. It moreover requires that personal data be adequate.

---

140  See further Recital 75 GDPR.

141  On the GDPR and risk management, see also Michèle Finck and Frank Pallas (n 30).

142  Sophie Stalla-Bourdillon and Alison Knight, 'Data Analytics and the GDPR: Friends or Foes? A Call for a Dynamic Approach to Data Protection Law' in Ronald Leenes et al (eds), *Data Protection and Privacy: The Internet of Bodies* (Hart 2018), 249.

143  Mireille Hildebrandt, 'Primitives of Legal Protection in the Era of Data-Driven Platforms', (2018) 2 *Georgetown Law Technology Review* 252, 267.

144  Datatilsynet, 'Artificial intelligence and privacy' (January 2018), 11 https://www.datatilsynet.no/globalassets/global/english/ai-and-privacy.pdf accessed 13 December 2019.

145  The French version of the GDPR indeed translates 'relevant' as 'pertinent'. The German language version of the GDPR indeed speaks of 'dem Zweck angemessen' - 'relevant for the purpose' in this context.

## 5.2 Adequacy

The requirements of relevance and adequacy are closely intertwined. Yet, there appears to be a nuance between both concepts. Whereas the relevance criterion has a purely limiting impact on data collection, in some circumstances adequacy may require that more data be processed. Indeed, omission of certain kinds of data can limit the usefulness and accuracy of a dataset and the analyses done on that dataset.[148] Minimisation is but one of various substantive requirements in Article 5 GDPR, others including fairness, transparency[149] and accuracy.[150] This provision ought to be interpreted holistically and its principles are to inform data minimisation and vice-versa. Using adequate data is indeed a means to ensure that a model is fair, transparent and accurate.

In some circumstances, adequacy will have a limiting effect on the quantity of data to be processed, such as where data that is inadequate in light of the purpose for which it is collected - as would be the case of the e-commerce website scenario above. However, in other circumstances, adequacy may require the processing of more data for data analysis to be fair and accurate. For example, it has been reported time and time again that many currently deployed models are inaccurate when it comes to certain demographic groups under-represented in training datasets. In such instances, processing more data could make the corresponding model more representative and thus help achieve the overarching requirements of fairness, transparency and accuracy.

Although formulated as part of the data 'minimisation' requirement, it hence seems that the adequacy requirement can actually mandate the processing of more rather than less personal data. The final requirement of the test under Article 5(1)(c), necessity, in contrast has a purely limiting scope.

## 5.3 Necessity

Finally, data should be 'limited' to what is necessary, meaning that controllers ought to identify the minimum amount of personal data needed to fulfil a purpose.[151] This is a somewhat stricter wording compared to the DPD, which required that personal data must not be 'excessive in relation to the purposes.'[152] As a consequence, anything exceeding the 'minimum' amount necessary will be an excessive processing, in breach of the data minimisation principle. For example, where the same results can be achieved through the processing of less personal data, or even of anonymous data, the processing of personal data can likely not be accepted as necessary.

It is worth noting that where there are multiple purposes, a data item can be necessary for one purpose but not for another, and the data controller can only process for the former. The necessity criterion is also crucial for the interpretation of Article 7(4) GDPR which requires that when assessing whether consent is freely given, 'utmost account shall be taken of whether, inter alia, the performance of a contract, including the provision of a service, is conditional on consent to the processing of personal data that is not necessary for the performance of that contract'.

These findings confirm that data minimisation continues to play a meaningful function in contemporary data processing practices. First and foremost, its relevance and necessity requirements impose limits on the quantity of data that can be processed. One could argue that an ill-intended controller could define a purpose in such a manner that the data they want to collect is "relevant" (returning to the above example, they could state that they explicitly want to provide astrological recommendations). Yet, the necessity and adequacy imperatives impose limits on the boundless collection of personal data even in cases like this. What is more, adequacy ensures that the right kind of data is collected, also in furtherance of other GDPR objectives such as adequacy and fairness. Finally, data minimisation requires controllers to preferentially process personal data that constitutes less risk for data subjects.

## 5.4 What Data Needs to be Minimised?

Article 5(1)(c) GDPR is generally interpreted as referring to the need to minimise the *quantity* of data that is processed. One may, however, also wonder whether the principle extends to other characteristics such as whether the data has been pseudonymised or whether it is special category data. This includes data on racial or ethnic origin, political opinions, religious or philosophical beliefs, trade union membership, genetic data, biometric data used for the purpose of uniquely identifying a natural person, health data and data concerning a natural person's sex life or sexual orientation.[153] What data actually qualifies as sensitive data at a time where sensitive characteristics can often be inferred from behaviour traces is a matter of ongoing debate.[154]

First, personal data should be anonymised or pseudonymized wherever possible. Whereas perfect anonymisation, which is hard to achieve, brings the processing outside the scope of the GDPR altogether, pseudonymisation can reduce the risk inherent to the processing. This position also seems to find support in the E-Privacy Directive, which speaks of the need 'of minimising the processing of personal data and of using anonymous or pseudonymous data where possible'.[155]

Second, Article 9 GDPR establishes a special regime for categories of data considered to reveal particularly sensitive information about indi-viduals. Article 9(1) establishes a general prohibition to process special category data (often also referred to as 'sensitive' data). In some circumstances, such data can nonetheless be processed where the data subject has provided explicit consent.[156] Under the GDPR, special category data can thus only be processed subject to conditions that are more burdensome for controllers than those arising under the ordinary regime. Explicit consent is the most relevant for profiling and personalisation systems. Whereas the concept of 'explicit consent' is not defined, it likely requires an oral or written affirmation of consent.[157] Sensitive data also ought not to be used to inform solely automated decisions that have legal or similarly significant effects on data subjects unless the data subject has explicitly consented or processing is necessary for reasons of substantial public interest.[158]

---

146  https://edpb.europa.eu/sites/edpb/files/files/file1/edpb_letter_out2020_0004_intveldalgorithms_en.pdf, 2-3.
147  International Working Group on Data Protection in Telecommunications, Working Paper on Privacy and Artificial Intelligence (n 6) 9.
148  Bart van der Sloot, 'From Data Minimization to Data Minimummization' in Bart Custers et al (eds) Discrimination and Privacy in the Information Society (Springer 2013) 274.
149  GDPR, art 5(1)(a).
150  GDPR, art 5(1)(d).
151  Information Commissioner's Office (n 68).
152  DPD, art 6(1)(c).
153  Article 9(1) GDPR.
154  See by way of example, Paul Quinn and Gianclaudio Malgieri, 'The Difficulty of Defining Sensitive Data – the Concept of Sensitive Data in the EU Data Protection Framework, *Brussels Privacy Hub Working Paper* (2020).
155  Directive 2002/58/EC of the European Parliament and of the Council of 12 July 2002 concerning the processing of personal data and the protection of privacy in the electronic communications sector (e-Privacy Directive) [2002] OJ L201/37, recital 9.

Personal data processed for profiling and personalisation often constitutes sensitive data,[159] such as when dating apps share their users' dating choices, information about drug use and ethnicity as well as precise geographical location with advertisers.[160] It is accordingly of pronounced practical importance for controllers of profiling and personalisation systems to determine whether their processing is caught by the GDPR's special regime.

Many agree that data minimisation not only entails an obligation to restrict the amount of data but also to keep sensitive data to a minimum. According to the Norwegian data protection authority, data minimisation 'stipulates proportionality' in intervening with a data subject's privacy.[161] This implies an obligation to restrict 'both the amount and the nature of the information used'.[162] As a consequence, pseudonymisation is encouraged as one measure limiting the identifiability of the data subject.[163] Zarsky concurs that minimisation 'relates to the scope and categories of data initially collected'.[164] This reflects that minimisation should not be seen as an isolated requirement but rather as a tool to interpret the entire GDPR. Indeed, the special regime created for special category data would substantiate the argument that the legislator intended for the processing of special category data to always be minimised.

Thus, data minimisation requires a limitation of sensitive data and at the same time, the latter is a frequent ingredient in personalisation systems. It is, however, doubtful that there is a legitimate basis for processing the data in light of the difficulty of achieving (explicit) consent. Supervisory authorities have for instance concluded that current consent requests in adtech do not comply with the requirements for explicit consent.[165] It thus appears that many profiling and personalisation systems are currently not compliant with the GDPR.

### 5.5 Current State

Despite the existence of appropriate computational techniques and empirical evidence suggesting that it might be possible to limit data in data-driven systems to a much larger extent, minimisation of user-generated, observational, and behavioral data does not appear to be a common practice. A study of software developers' approaches to data minimisation revealed that practitioner practices differ both in terms of protocols and tools,[166] highlighting the need for more specific implementation guidelines.  The study, however, did not cover approaches to data minimisation in data-driven systems and, to the best of our knowledge, no such study exists.

Guidelines for implementing data minimisation have been issued by both the British[167] and Norwegain[168] data protection authorities. These guidelines suggest techniques that could be used to minimise data such as investigation of learning curves—a technique which ties minimisation to performance, similarly to one of the recently proposed formal minimisation interpretations.[169] Still, the suggestions do not go into lower-level operational details. Several open computational questions are a potential reason why more detailed guidelines for performance-based data minimisation are missing. The next section presentes these questions in detail.

### 5.6 Computational Challenges

Minimising data for the purpose of improving a system's performance faces a number of obstacles. The first and foremost challenge lies in determining whether and which data improves results. The state-of-the-art computing knowledge does not provide off-the-shelf answers. The closest relevant line of work aims at quantifying the impact of individual data points in training sets on the accuracy of machine learning models trained using those sets.[170] Such data points often correspond to individual persons and are composed of multiple pieces of information (features). To the best of our knowledge, methods that would quantify which individual pieces of data about an individual data subject are necessary to improve a personalized service (for the individual or globally) or quantify the improvement itself, are missing. The problem in fact poses a number of computational and research challenges.

Furthermore, determining whether a piece of personal data should be minimized out will be a form of prediction about future system performance. As such, these predictions can reasonably be expected to be inaccurate and a question remains what level of inaccuracy would be acceptable for this form of data minimisation to be deemed practically viable.

Beyond the prediction accuracy, the determination of how much and what personal data is to be kept for a given performance purpose, depends on a number of factors. Those include the prediction method itself, the underlying service algorithm, existing user data, as well as the entirety of other data at the disposal of the service provider. Out of those factors, two merit special attention. First, data minimisation outcomes will largely depend on the underlying algorithm. Advanced systems employing complex models that need to learn many parameters require enough data to function properly.

---

156  GDPR, art 9(2)(a).
157  Information Commissioner's Office (n 68).
158  GDPR, art 22(4).
159  Sandra Wachter, 'Affinity Profiling and Discrimination by Association in Online Behavioural Advertising' (2020) *Berkely Technology Law Journal* (forthcoming).
160  Natasha Singer and Aaron Krolik, 'Grindr and OkCupid Spread Personal Details, Study Says' New York Times (13 January 2020) https://www.nytimes.com/2020/01/13/business/grindr-apps-dating-data-tracking.html accessed 31 January 2020.
161  Datatilsynet (n 144).
162  Ibid.
163  Ibid.
164  Tal Zarsky, 'Incompatible: GDPR in the age of big data' (2017) 47 *Seton Hall Law Review* 995, 1009.
165  For an overview, see Information Commissioner's Office, (n 127).
166  Awanthika Senarath and Nalin Asanka Gamagedara Arachchilage, 'Understanding Software Developers' Approach towards Implementing Data Minimization' [2018] arXiv:1808.01479 [cs] http://arxiv.org/abs/1808.01479 accessed 2 December 2020.
167  Information Commissioner's Office (n 68).
168  Datatilsynet (n 144).
169  Asia J Biega and others (n 16).0
170  See, for example, Richard Chow and others, 'Differential Data Analysis for Recommender Systems' [2013] *Proceedings of the 7th ACM conference on Recommender systems* 323; Amirata Ghorbani and James Zou, 'Data Shapley: Equitable Valuation of Data for Machine Learning' (2019) 97 *Proceedings of the 36th International Conference on Machine Learning* 2242.

Examples include deep learning methods, whose recent reemergence has been made possible by the availability of vast image datasets[171], or natural language understanding methods that benefited from comprehensive Web corpora[172]. Thus, by developing and implementing state-of-the-art solutions, service providers might be able to collect more personal data. On the other hand, infusing vanilla models with domain-specific knowledge might allow for better performance with smaller models, and thus less need for data collection[173]. These observations further lead to a number of questions of economic nature. If less data can be collected with custom models, will companies who can afford an internal research unit be able to minimise data better? If bigger models grant a data processor the right to collect more data, will companies who can afford the costly infrastructure necessary to operate those models be allowed to collect disproportionately more data?

The second factor influencing data minimisation that is worth highlighting is the complex balance and interdependence of the data of different users and the system performance for those users. It might be tempting to think of the system data as static when considering which pieces of an individual's personal data to minimise out. However, the data that is minimized out for a given individual might constitute the system training data for other individuals. Thus, minimisation of data for a single user will also influence the performance of the system for other users. The need for a global, systemic approach that at the same time works for each individual separately, makes reasoning about data minimisation ever so complex and raises the question of whether we should acknowledge minimisation dependencies analogous to privacy dependencies.[174]

Last but not least, taking a user's perspective, it is important to recognise that different people might have different *personal purposes* when using a service in a seemingly same way. For instance, a user might generate a movie rating purely to give the provider the information needed for personalized movie recommendations, in which case the performance-based minimisation appears appropriate. Another user, however, might generate movie ratings for personal archiving purposes—to store a log of movies they have seen. In this case, the storage of all personal ratings appears appropriate. As a result, it might be necessary to not only model the purpose of data collection by the service provider, but also the *purpose of data generation* by the user.

The need to model user personal purposes leads to two challenges. First, it is rather difficult to infer user intent from their behavior. Recognising people's search intents from text queries, for instance, continues to be an active research problem in information retrieval.[175] Second, users might engage with technology products in originally unanticipated ways. Some people, for example, use email—developed primarily as a communication tool—as a task management and archiving system. Studies show that some users send themselves emails with todos, reminders, or files to archive.[176]

Inaccurate detection of user personal purposes might lead to both under- and over-minimisation of data, depending on the context. If personal purposes were recognized correctly, a system might adapt its minimisation strategies to a particular intent. For instance, if a user generated certain data points purely for archiving purposes, a system might store the data separately from all other user data and not use it for any other purpose, such as improving personalisation. Data generated for personalisation, on the other hand, might be minimised based on performance goals.

## 5.7   Trade-Offs Inherent to Data Minimisation

Just as purpose limitation, data minimisation presents numerous trade-offs. We again see the explicit and acknowledged trade-off between the risks and benefits of personal data usage. Data minimisation is essentially a risk-management tool which minimises risks by limiting the quantities and categories of data that can be lawfully processed.

The practical application of data minimisation, however, also results in numerous unexpected trade-offs. First, we have demonstrated that the determination of whether a piece of personal data should be minimised is a form of prediction about future system performance—which may be inaccurate. Controllers with inaccurate performance prediction algorithms might be rewarded by seemingly legitimate increased personal data collection. Second, data minimisation may drive the data controller to employ algorithms which are less robust to minimisation to be able to collect more data. In case such algorithms also offer worse performance, end users would end up penalised with both increased data collection and decreased satisfaction.

Finally, our analysis has revealed that data minimisation has collective rather than purely individual consequences. Minimisation of a user's data will impact system performance for other users and it will be important to understand the impact of individual subject's choices and preferences on the collective system dynamics.

## 6.   Purpose Limitation and Data Minimisation Highlight Important Trade-Offs in Data Protection Law

Beyond the trade-offs inherent to purpose limitation and data minimisation, our research has further exemplified a number of other trade-offs inherent to data protection law at large.

## 6.1   The Generality of Legal Principles and The Need for Computationally Operational Interpretations

In order to comply with purpose limitation and data minimisation, computer scientists need measurable definitions of those principles as well as specific implementation guidelines. Only with precise mathematical definitions can algorithms determine which data to retain and which to discard, or predict whether data will improve services as it is collected. This need stands in tension with the GDPR as a general, principles-based and technology-neutral legal framework. Indeed, the GDPR and its implementing guidance do not provide any concrete indications to computing practitioners as to how to practi-cally and concretely implement their legal requirements. As a result, currently it is difficult to determine whether a given computation adheres to the pre-defined purpose or whether collected data is adequate, relevant and necessary. Perhaps unsurprisingly, practitioners apply various, often inconsistent, approaches to minimisation.[177]

---

171  Jia Deng and others, 'ImageNet: A Large-Scale Hierarchical Image Database' [2009] *2009 IEEE Conference on Computer Vision and Pattern Recognition* 248.

172  Alon Halevy, Peter Norvig and Fernando Pereira (n 11).

173  For instance, smaller custom-tailored models have been shown to outperform vanilla language models in dialogue systems. See: Matthew Henderson and others, 'ConveRT: Efficient and Accurate Conversational Representations from Transformers' [2020] *Findings of the Association for Computational Linguistics: EMNLP 2020* 2161.

174  Solon Barocas and Karen Levy, 'Privacy Dependencies' (2020) 95 *Washington Law Review* 555.

175  For example: Hamed Zamani and others, 'Generating Clarifying Questions for Information Retrieval' [2020] *Proceedings of The Web Conference 2020* 418; Bernard J Jansen, Danielle L Booth and Amanda Spink, 'Determining the User Intent of Web Search Engine Queries' [2007] *Proceedings of the 16th international conference on World Wide Web* 1149.

176  Horatiu Bota and others, 'Self-Es: The Role of Emails-to-Self in Personal Information Management' [2017] *Proceedings of the 2017 Conference on Conference Human Information Interaction and Retrieval* 205.

This trade-off between the value of general legal principles and the practical need for concrete interpretations could be addressed from both the legal and computational ends. On the legal side, it might be possible to develop more specific guidance. The European Data Protection Board would have to issue concrete overall guidance which might then be rendered more concrete when implemented at a firm level (requiring collaborations between technical and legal experts).

On the computational side, researchers could develop new technical implementation proposals which then could be evaluated by legal experts. Many algorithmic techniques that will likely be useful for automating data minimisation already exist (including, for instance, feature selection, outlier detection, analysis of learning curves, or active learning), even though they need to be adapted to adequate interpretations of data minimisation and purpose limitation. A recent line of work in computer science offers a glimpse of how we might attempt to interpret the principles in personalisation and profiling systems. Biega et al. proposed to interpret the purpose of data collection in data-driven systems as improvement in system performance metrics.[178] Shanmugam et al. proposed a framework for data minimisation based on algorithmic performance curves,[179] while Goldsteen et al. proposed a framework leveraging data anonymisation techniques.[180]

Our analysis has, however, highlighted the difficulties of automating legal compliance. Indeed, it may well be that in many scenarios measuring compliance with purpose limitation and data minimisation is simply too burdensome and costly. Similar difficulties can also be observed in respect of the computational implementation of another core GDPR principle, fairness.[181] Indeed, recent interdisciplinary work has highlighted that the legal prohibitions of certain kinds of discrimination (conventionally considered to be at the core of fairness) are too contextual, reliant on intuition and open to judicial interpretation to be automated. Thus, it is likely that many of the computational implementations of fairness, including "fairness toolkits" are unable to adequately reflect legal requirements.[182] Whereas this paper will not elaborate on these discussions in further detail, future work should more closely evaluate both the desirability and necessity of automating legal compliance. If automating compliance requires a fundamental change in law's contextual nature, discussions ought to be had about the implications and desirability of such changes.

The effort to translate the principles of purpose limitation and data minimisation into practice in data-driven systems will likely require an extensive dialogue between the legal and computational communities to determine which interpretations are viable both legally and computationally, much like the dialogue that has happened for the principles of antidiscrimination or fairness, and which has spun a large body of research in both communities.

### 6.2 The Unacknowledged Trade-Offs Between Various GDPR Principles

The above analysis has illustrated that the legal data minimisation principle requires that data usage is kept to the necessary minimum. Data minimisation hence encourages a restrictive processing of data, assuming that such restricted processing is preferred from a data protection perspective. It is important to acknowledge, however, that there is a trade-off between restrictive data usage and other GDPR objectives, such as fairness. Indeed, complying with fairness (which, it is important to point out, remains under-defined from a legal perspective) may require processing more data. Recent empirical studies in the domain of recommender systems have suggested that limiting data might have disparate consequences for individuals[183] and user groups,[184] while minimisation of sensitive features (such as gender) may moreover limit our ability to audit fairness.[185]

Furthermore, Article 5 (1)(d) GDPR requires that personal data be "accurate and, where necessary, kept up to date; every reasonable step must be taken to ensure that personal data that are inaccurate, having regard to the purposes for which they are processed, are erased or rectified without delay". This accuracy requirement may also compel the data controller to collect more data. For instance, personalisation profiles consisting of product ratings or search queries may become obsolete if data collection stops because of the data minimisation requirement, yet the interests and preferences of data subjects change. In this context, it might seem necessary to continuously collect new data while focusing minimisation on the old data. In fact, Google introduced such 3-18 months auto-deletion of search and location data in 2019,[186] and made it the default setting for new users in 2020.[187]

The fact that data minimisation promotes reliance on restrictive quantities of data whereas other GDPR principles such as fairness and accuracy will sometimes require the collection of additional data raises the question of how these objectives ought to be reconciled in practice. All these requirements constitute core data protection principles enshrined in Article 5 GDPR—which does not establish a hierarchy among its various requirements. As such, it cannot be concluded that data minimisation is a superior objective compared to fairness or vice versa. Thus, in practice, computer scientists must make sure that data minimisation as well as fairness and accuracy are equally respected.

Data minimisation by itself already incorporates leeway for such balancing of principles through its three requirements of relevance, adequacy and necessity. The collection of further data to comply with the fairness or accuracy requirements can, depending on the circumstances, be considered to be relevant, adequate and necessary. Indeed, our analysis above confirmed that the adequacy requirement can itself be read as requiring the collection of more data to comply with considerations such as fairness. From a legal perspective, however, what is relevant, adequate and necessary should be determined

---

on a case-by-case basis, taking into account contextual factors.

## 6.3 Data Subject Rights and the Economic and Environmental Costs of Enforcing Them

Our analysis has highlighted that personal data can be processed for purposes exceeding the initially defined purposes with a data subject's consent. At the same time, the GDPR provides that whenever personal data processing is legitimised through consent, the data subject subsequently has the right to withdraw his or her consent at any time.[188] Whereas the withdrawal of consent does not affect the lawfulness of past personal data processing, it prohibits the data controller from continuing to process that personal data in the future. Machine learning models are trained on already collected personal data but are employed to make new inferences—a form of personal data processing. Thus, withdrawal of consent to process personal data encoded in a model requires deconstructing the model. How to efficiently remove individual data points from trained machine learning models is a subject of active research.[189] Currently, in cases of consent withdrawal, models might have to be continually retrained, yielding computational and thus also environmental costs. How to balance enforcement of individual data rights *vis-à-vis* environmental costs is a pertinent question.

## 6.4 The Cost of Compliance and the Unlikelihood of Enforcement

Ultimately, the success of policies, including data protection, hinges on their practical implementation. Unless there is adequate enforcement of related provisions, it is doubtful whether addresses have sufficient incentives to enforce related legal requirements, particularly if such implementation is costly. Compliance with data protection law is indeed costly. It requires data controllers to contract related expertise as well as carefully designing their technical and organizational structures. Perhaps most significantly, it also prevents them from pursuing forms of data analysis that may be attractive from a business perspective yet risky in terms of violating data protection law.[190] As such, it is important that data protection law is properly implemented. At present data controllers will rationally make a trade-off between the economic benefits of unconstrained usage of personal data and the potential yet very unlikely economic cost of data protection enforcement.

Recent years have, however, underlined that the enforcement of the GDPR is riddled with hurdles, such as the uneven geographical distribution of relevant competence (on the basis of a company's seat in the EU) or the fact that data protection authorities have insufficient means to meaningfully police compliance with the Regulation.[191] Indeed, even though Article 52(4) GDPR requires that supervisory authorities have the required technical resources, evidence is mounting that these resources are currently insufficient.[192]

Another factor that is highly relevant with respect to purpose limitation and data minimisation relates to the technical difficulties of verifying compliance. Indeed, whereas anyone, including supervisory authorities, can in most circumstances consult a data controller's data protection policies to read or automatically analyse[193] how the purposes are defined and what data is acknowledged to be processed, verifying whether these statements are honored in practice is an entirely different matter. Determining whether individual pieces of data are necessary for personalisation might be computationally difficult, and in general more research would be necessary to establish what is the form of evidence that data controllers should produce to prove compliance. Furthermore, measures of minimisation could be gamed. Since improvements in the results are often functions of complex interactions between individual pieces of data, it is feasible to imagine a data collection mechanism that requests a large set of data, where seemingly all items are necessary, even though there exists a smaller set of data yielding a similar performance that could have been collected instead. While continual reassessment of whether existing data is necessary is mandated, it might be impossible to determine whether the retained data in fact is the minimum data.

## 7. Outlook

This paper has shown that, despite what has been suggested by many commentators, purpose limitation and data minimisation remain feasible albeit challenging in the context of data-driven personalisation, profiling and decision-making systems. At the same time, they force data controllers to make many, oftentimes unacknowledged, trade-offs. While the longer-term research problems await their solutions, practitioners might employ a variety of organizational and technical best practices as well as off-the-shelf tools that minimise data even if not explicitly developed for minimisation purposes.

## 7.1 Short-term Practitioner Guidelines

Even though the implementation of purpose limitation and data minimisation in the context of data-driven systems bears a considerable research agenda, practitioners might consider implementing a range of existing solutions and best practices that contribute toward data minimisation. The first organizational best practice is for employers and employees to create and cultivate a mindset of reflecting on the purposes of data they collect, continuously considering whether data should be collected and when it should be deleted. To quantify the importance of different pieces of data, practitioners can use off-the-shelf solutions for machine learning models, including feature selection, data influence estimation, or data valuation. At the data collection time, techniques such as active learning would allow data processors to prioritize which data is the most important for a model's quality. Data can also be minimised through simpler heuristics, such as selection of representative random samples, or selection of data specifying certain domain-specific quality criteria, or retaining of the most recent data only. For instance, in the context of product recommendations, a data controller might retain only the most recent product ratings generated by a user, only the ratings for the most popular products, or only the ratings with the highest or lowest values. Finally, minimisation should be employed not only at the data collection time, but also continually reapplied to existing data stores.

---

188   Article 7(3) GDPR.
189   Some of the recently proposed approaches include: Antonio Ginart and others (n 138); Lucas Bourtoule and others, 'Machine Unlearning' [2020] arXiv:1912.03817 [cs] http://arxiv.org/abs/1912.03817 accessed 3 December 2020; Sanjam Garg, Shafi Goldwasser and Prashant Nalini Vasudevan, 'Formalizing Data Deletion in the Context of the Right to Be Forgotten' (2020) 12106 *Advances in Cryptology – EUROCRYPT 2020* 373; Chuan Guo and others, 'Certified Data Removal from Machine Learning Models' (2020) 119 *Proceedings of the 37th International Conference on Machine Learning* 3832.
190   See also Articles 5(2) and 25(1) GDPR.
191   Derek Scally, 'German Regulator Says Irish Data Protection Commission Is Being "Overwhelmed"' (*The Irish Times*, 3 Feb 2020) https://www.irishtimes.com/business/financial-services/german-regulator-says-irish-data-protection-commission-is-being-overwhelmed-1.4159494 accessed 3 December 2020.
192   https://brave.com/wp-content/uploads/2020/04/Brave-2020-DPA-Report.pdf
193   Various natural language processing (NLP) techniques have been proposed to automatically extract or align policy statements; see, eg Shomir Wilson and others (n 79). Further techniques include automated question answering, allowing readers to obtain concise answers to their questions about a given verbose policy, see Abhilasha Ravichander and others (n 79).

## 7.2 Long-term Research

To be technically implementable in the context of data-driven systems, purpose limitation and data minimisation will likely need to follow a similar research trajectory as that of work in algorithmic fairness. As demonstrated throughout the paper, we need mathematical interpretations of the principles, decision rules for deciding which pieces of data are necessary and which should be discarded, machine learning models that could automate compliance, and quantitative data analyses for understanding how the implementation of those principles might influence the quality and functioning of online ecosystems. We need standardisation of data processing purposes to ensure their specificity as well as an understanding of how different purposes relate to each other to reason about their compatibility. We lack an in-depth understanding of what value people associate with different types of data in different contexts. We moreover should design appropriate transparent mechanisms for collecting informed data processing consent as well as technical means of removing data from existing models and infrastructures once a purpose has been fulfilled or a user withdraws their consent. We need auditing methods that could establish compliance with the purpose limitation and data minimisation requirements. Finally, we need to establish normatively, legally, and technically, how to balance data minimisation with other GDPR requirements—such as fairness or accuracy—which might be at odds with the minimisation principle.

Yet, as the work on algorithmic fairness has previously exemplified, it is difficult to bridge terminological and substantive gaps between disciplines.[194] One may wonder whether and how legal principles, particularly broad principles purposefully kept vague to enable contextual interpretation, can be translated into computer code. Without doubt, this is a question at the heart of digitalisation that requires more engagement from multiple disciplines in the years to come.

## Acknowledgements

We thank Solon Barocas for discussions that helped ideate the interdisciplinary direction of this work, as well as Hal Daumé III, Fernando Diaz, Peter Potash, Samira Shabanian, and Divya Shanmugam for fruitful discussions on computational interpretations of purpose limitation and data minimisation. We are also grateful to Kai Ebert for excellent research assistance.

---